\documentclass[twocolumn,trackchanges]{aastex701}

\usepackage{CJK}
\usepackage{booktabs} 
\usepackage{graphicx}
\usepackage{stackengine}
\begin{document}

\title{A Systematic Search for Broad Absorption Line Outflows in Changing-Look Active Galactic Nuclei}

\author[orcid=0009-0008-4874-6502]{Bin Duan}
\email[hide]{dbdb@mail.ustc.edu.cn}  
\affiliation{Department of Astronomy, University of Science and Technology of China, Hefei 230026, People’s Republic of China}
\affiliation{School of Astronomy and Space Science, University of Science and Technology of China, Hefei 230026, People’s Republic of China} 

\author[orcid=0000-0003-3667-1060]
{Zhicheng He} \email[]{zcho@ustc.edu.cn}  
\affiliation{Department of Astronomy, University of Science and Technology of China, Hefei 230026, People’s Republic of China}
\affiliation{School of Astronomy and Space Science, University of Science and Technology of China, Hefei 230026, People’s Republic of China} 

\author{Guilin Liu}
\affiliation{Department of Astronomy, University of Science and Technology of China, Hefei 230026, People’s Republic of China}
\affiliation{School of Astronomy and Space Science, University of Science and Technology of China, Hefei 230026, People’s Republic of China}
\email[]{glliu@ustc.edu.cn}  

\author{Yu Rong}
\affiliation{Department of Astronomy, University of Science and Technology of China, Hefei 230026, People’s Republic of China}
\affiliation{School of Astronomy and Space Science, University of Science and Technology of China, Hefei 230026, People’s Republic of China}
\email[]{rongyua@ustc.edu.cn}  

\author{Zhenyi Cai}
\affiliation{Department of Astronomy, University of Science and Technology of China, Hefei 230026, People’s Republic of China}
\affiliation{School of Astronomy and Space Science, University of Science and Technology of China, Hefei 230026, People’s Republic of China}
\email[]{zcai@ustc.edu.cn}

\correspondingauthor{Zhicheng He}
\email{zcho@ustc.edu.cn}


\begin{abstract}

Changing-look active galactic nuclei (CLAGNs) provide a probe of the relationship between accretion-state transitions and AGN outflows. We systematically searched for broad absorption line (BAL) signatures in a well-defined sample of 370 optically selected CLAGNs and identified one LoBAL source, SDSS J160139.10+412529.3, corresponding to an observed LoBAL fraction of $0.27\%_{-0.17\%}^{+0.44\%}$. The available spectra do not permit an unambiguous ionic identification of the low-ionization absorption trough, which may arise from \ion{Mg}{2} or \ion{Fe}{2}. Multi-epoch spectroscopy from SDSS and DESI, together with long-term photometric monitoring, shows a $\sim$2 mag decline in the $g$ band over two decades, accompanied by substantial weakening of both the \ion{Mg}{2}/\ion{Fe}{2} BAL feature and the broad emission lines. The single LoBAL detection in this sample suggests that detectable low-ionization BAL outflows may be uncommon among optically selected CLAGNs, although observational selection effects may also contribute. The observed LoBAL fraction is about a factor of two lower than that in ordinary AGNs with similar Eddington ratios, but the difference is not statistically significant. Larger samples and ultraviolet spectroscopy are required to establish the incidence of BAL outflows among CLAGNs and to determine the ionic origin of the absorption in SDSS J160139.10+412529.3.
\end{abstract}

\section{Introduction}

Broad absorption line (BAL) active galactic nuclei (AGNs) and changing-look AGNs (CLAGNs) represent two of the most remarkable manifestations of AGN activity. BAL AGNs exhibit broad and blueshifted absorption troughs that trace high-velocity outflows, which are widely regarded as an important channel of AGN feedback. In contrast, CLAGNs undergo dramatic spectral transformations characterized by the appearance or disappearance of broad emission lines accompanied by large-amplitude continuum variability \citep[e.g.,][]{Ricci2023,guoHigh2020,guo2024}. Both phenomena provide valuable insights into the interaction between supermassive black holes and their surrounding environments.

Previous studies have suggested that the occurrence of BALs may depend on both viewing geometry \citep[e.g.,][]{Elvis2000} and the physical conditions of the central engine \citep[e.g.,][]{Hazard1984,Zhang2010,Vivek2025}. Likewise, changing-look transitions are thought to be associated with substantial changes in the accretion flow \citep[e.g.,][]{Graham:2020nz,MacLeod2016,MacLeod2019,sheng2017,Kokubo:2020kr} and may also be influenced by geometric effects in some systems \citep[e.g.,][]{Mereghetti:2021aa,Burtscher16}. Therefore, both BAL AGNs and CLAGNs are likely drawn from specific regions of AGN parameter space defined by orientation, accretion state, or a combination of both.

Previous spectroscopic studies have identified and characterized samples of CLAGNs primarily through changes in their broad emission lines \citep[e.g.,][]{MacLeod2016,MacLeod2019,Yang2018,Hon2022,Temple2022,guo2024,Guo2025}. BAL-related spectral changes have also been considered in individual objects or as a possible complication in identifying changing-look transitions. In addition, the connection between BAL variability and changes in the ionizing continuum has been extensively investigated \citep[e.g.,][]{Wang2015,He2017,He2025}. These studies provide important context for a possible connection between BAL and changing-look phenomena, although they were not designed to measure the incidence of BALs within an identified CLAGN population.

The degree of overlap between these two populations provides a direct observational test of their underlying models. If BAL AGNs and CLAGNs occupy similar ranges of viewing angle and accretion state, a substantial fraction of CLAGNs would be expected to exhibit BAL features. Conversely, if the physical or geometric conditions required for the two phenomena differ significantly, their coexistence should be intrinsically rare. Measuring the incidence of BALs within a defined CLAGN population can therefore constrain the physical conditions associated with BAL outflows and changing-look transitions.

In this work, we perform a systematic search for BAL signatures in a well-defined sample of $370$ optically selected CLAGNs \citep{Guo2025}. We identify one LoBAL source, \object{SDSS J160139.10+412529.3}, corresponding to an observed LoBAL fraction of $0.27\%_{-0.17\%}^{+0.44\%}$. We further investigate its multi-epoch spectral evolution, quantify the LoBAL incidence in the CLAGN sample, and discuss both its possible physical implications and the observational selection effects that may influence the measured fraction.

\section{Sample and BAL Identification} \label{sec:data}

We utilize the sample of 561 changing-look AGNs (CLAGNs) from \citet{Guo2025}, selected from 49,633 cross-matched spectral pairs between SDSS DR16/DR16Q and DESI DR1. \citet{Guo2025} performed spectral decomposition on $4\,\text{\AA}\,\text{pixel}^{-1}$ rebinned spectra, accounting for host-galaxy starlight, the AGN power-law continuum, Fe~\textsc{ii} emission, and multi-component emission lines. To mitigate systematic uncertainties across disparate epochs and instruments, the catalog implements a three-stage calibration: internal flux normalization via the \ion{O}{3}~$\lambda5007$ line (within 20\%), pseudophotometry validation against CRTS/PS1/ZTF light curves ($<0.5\,\text{mag}$), and corrections for SDSS ($2''/3''$) and DESI ($1\farcs5$) fiber aperture effects. A broad-line flux ratio $R > 1.5$ and a significance threshold of $\text{Max}(N\sigma) > 3$ were required in constructing the CLAGN sample.

Starting from the 561 CLAGNs, we retrieved the 
corresponding optical spectra from SDSS (including the BOSS survey) \citep{SDSSdr7,BOSS,BOSS_SDSS3} and DESI DR1 \citep{DESIdr1} 
to perform a systematic search for Broad Absorption Line (BAL) features. To 
ensure that the \ion{Mg}{2}~$\lambda\lambda2796, 2803$ doublet remains within the 
spectral coverage of both SDSS/BOSS and DESI, we restricted our sample to 
sources with redshifts $z \ge 0.33$, yielding a subsample of 370 objects for 
further analysis. 

\begin{figure*}[t!]
\centering
\includegraphics[width=0.98\textwidth]{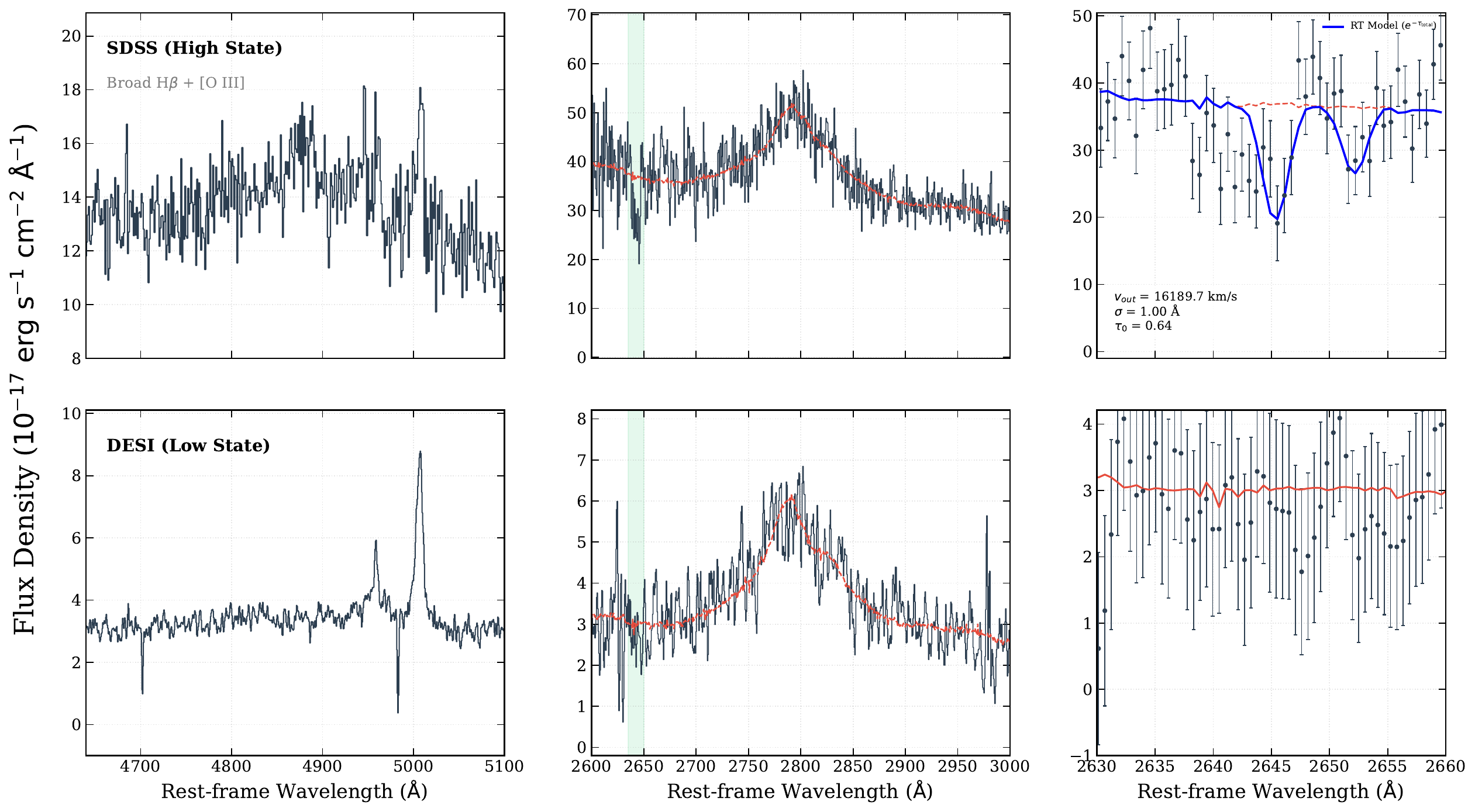}
\caption{Multi-epoch evolution of the broad emission lines and the low-ionization broad absorption feature during the changing-look event of J160139.10+412529.3. The upper and lower rows display the spectra observed during the SDSS high state and the later DESI lower state, respectively.
\textit{Left column}: Zoom-in view of the \text{H}$\beta$--[\ion{O}{3}]~$\lambda\lambda4959,5007$ region ($4640$--$5100$~\text{\AA}). The broad \text{H}$\beta$ emission line becomes substantially weaker from the high state to the lower state.
\textit{Middle column}: Overview of the rest-frame $2600$--$3000$~\text{\AA} spectral region. The dashed red lines denote the intrinsic NMF continuum, and the green shaded region marks the position of the BAL trough.
\textit{Right column}: Kinematic and component fitting of the possible \ion{Mg}{2}/\ion{Fe}{2} absorption complex ($2630$--$2660$~\text{\AA}). For the SDSS high state, the data points are shown together with the best-fit absorption model (blue line). Under the \ion{Mg}{2} interpretation, the model is decomposed into the \ion{Mg}{2}~$\lambda2796$ and $\lambda2803$ components. In the later DESI spectrum, the absorption feature is substantially weaker than in the SDSS high-state spectrum.}
\label{fig:1}
\end{figure*}

The unabsorbed (intrinsic) continuum and emission-line profiles were reconstructed 
using Non-negative Matrix Factorization (NMF) with 10 principal basis components \citep{Zhu2016}. 
These components were derived from a high-quality training set of unabsorbed 
quasar spectra selected from the SDSS DR12 catalog, restricted to $1 < z < 2.4$ 
and a signal-to-noise ratio ($\text{SNR}$) $> 10$. To reduce the influence of absorption features and outlying pixels on the reconstruction, we 
employed an iterative $3\sigma$-clipping procedure: in each of the three 
iterations, pixels deviating from the model by more than $3\sigma$ were masked. 
Following this reconstruction, we quantified the absorption strength by 
calculating the Absorption Index (AI) and the Balnicity Index (BI) within the 
identified troughs. Finally, each spectrum with AI or BI larger than zero underwent manual visual inspection to 
assess the absorption features and exclude artifacts. This 
screening process identified one LoBAL source that satisfied all 
selection criteria for detailed study (Figure~\ref{fig:1}).

\section{Detailed Multi-Epoch Spectral Processing} \label{app:spectral_details}

In this section, we present the technical details regarding the multi-epoch spectral alignment, flux calibration, and individual multi-component profile fittings performed on J160139.10+412529.3.

\begin{figure*}[htbp]
\centering
\newcommand{\stacklbl}[2]{\stackinset{l}{5pt}{t}{5pt}{\textbf{#1}}{#2}}
\begin{minipage}[b]{0.48\textwidth}
\centering
\stacklbl{(a)}{\includegraphics[width=\textwidth]{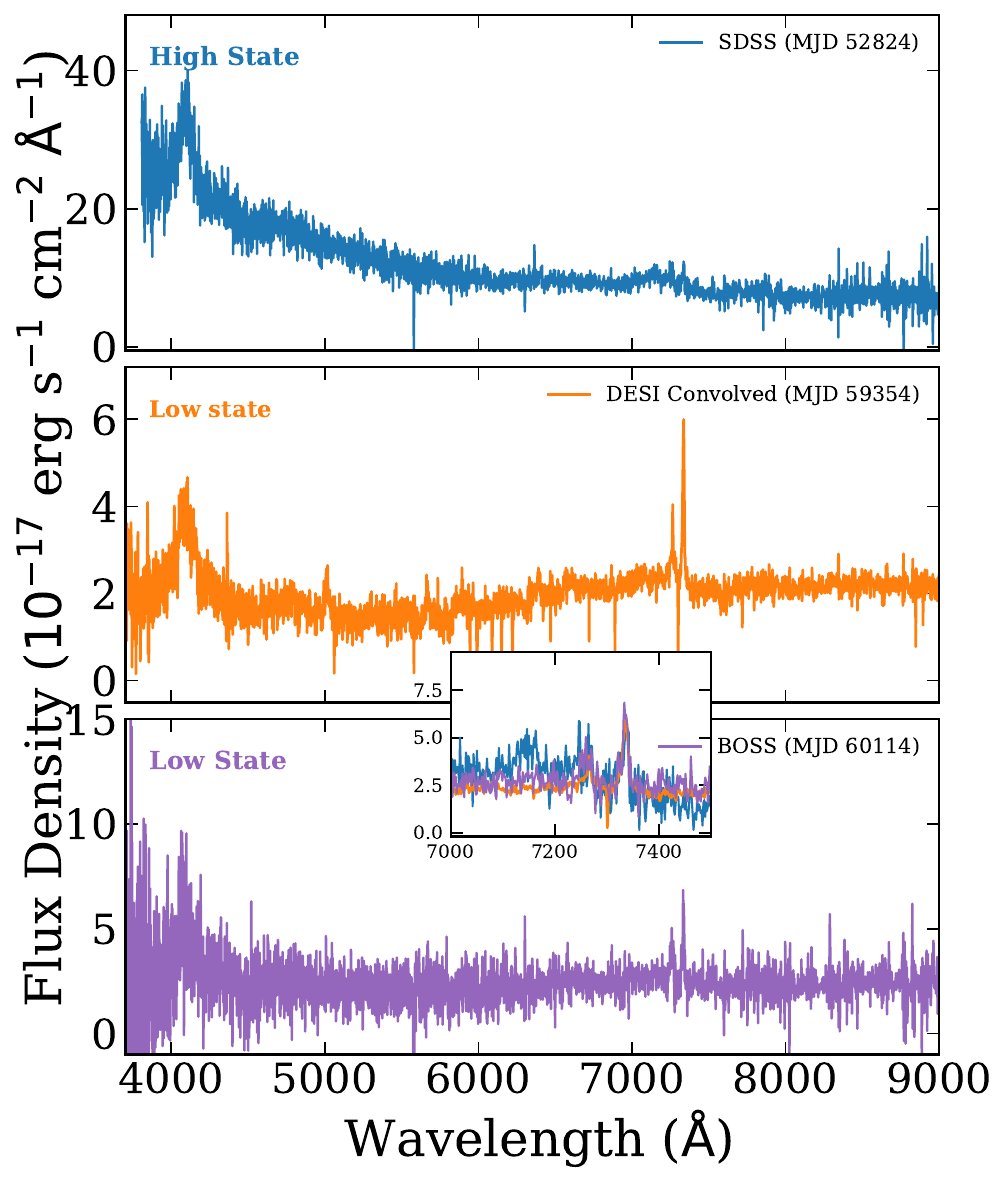}}
\end{minipage}
\hfill
\begin{minipage}[b]{0.48\textwidth}
\centering
\stacklbl{(b)}{\includegraphics[width=\textwidth]{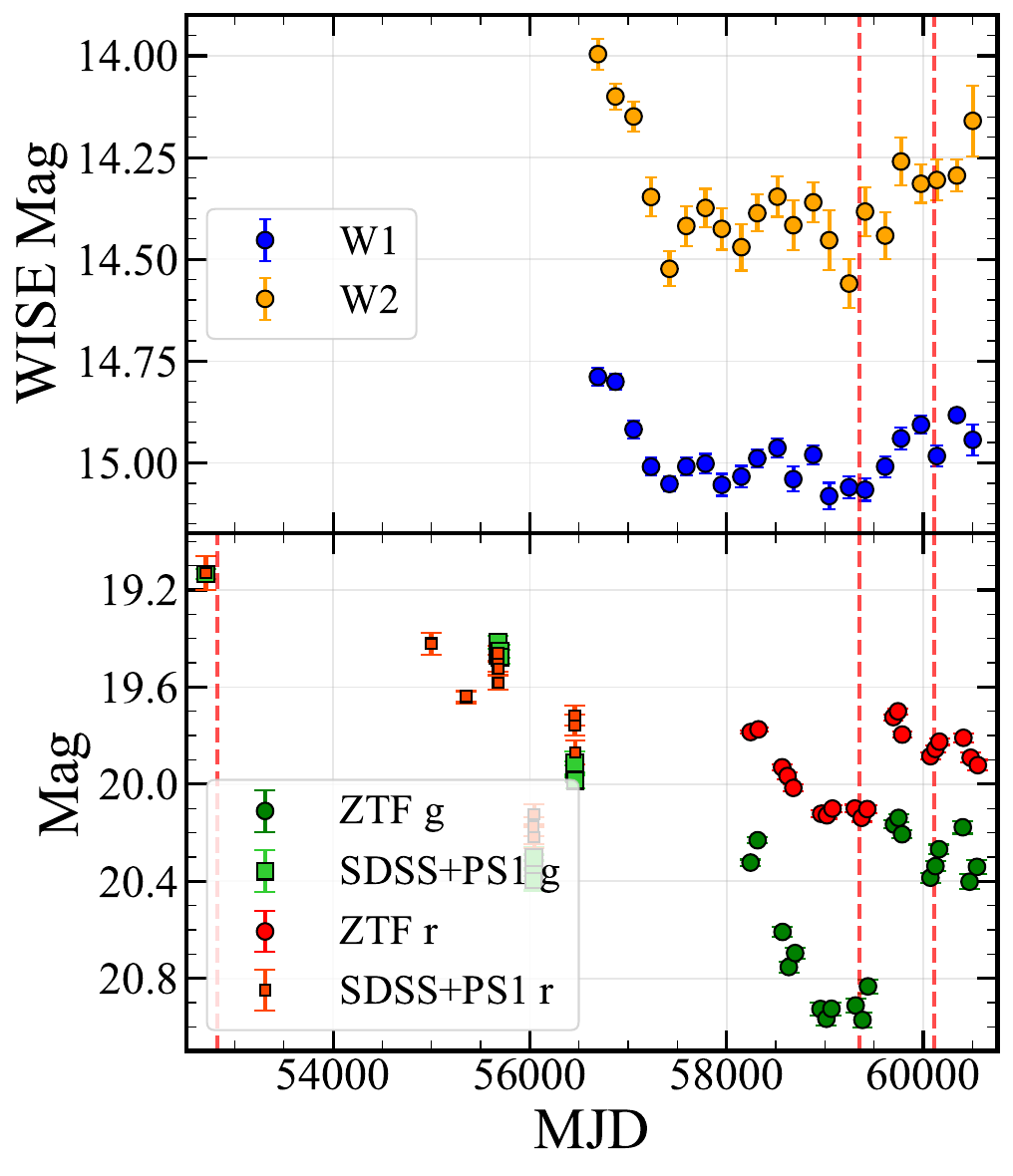}}
\end{minipage}
\caption{(a) Left panels: Multi-epoch rest-frame spectral evolution and changing-look behaviors of the target across distinct accretion states. From top to bottom, the high-state SDSS (MJD~52824; dark sky blue), convolved intermediate-state DESI (MJD~59354; vermillion), and low-state BOSS (MJD~60114; purple) spectra are plotted chronologically. The inset shows the alignment of the [\ion{O}{3}]~$\lambda5007$ narrow profiles after flux normalization. (b) Right panels: Long-term multi-band photometric light curves. The top sub-panel illustrates the mid-infrared \textit{NEOWISE} $W1$ (blue) and $W2$ (orange) magnitudes. The bottom sub-panel presents the optical $g$- and $r$-band measurements from SDSS, Pan-STARRS1, and ZTF, as indicated by the symbols in the legend. The three vertical red dashed lines mark the spectroscopic epochs of SDSS (MJD~52824), DESI (MJD~59354), and SDSS-V/BOSS (MJD~60114).}
\label{fig:2}
\end{figure*}

To eliminate instrumental systematic effects and resolution mismatches among the SDSS, DESI, and SDSS-V/BOSS spectrographs, the higher-resolution DESI spectrum was convolved with a tailored Gaussian kernel to match the lower-resolution linespread functions of the other two epochs. For the internal cross-instrument flux recalibration, we anchored the narrow [\ion{O}{3}]~$\lambda5007$ line flux in all spectra to the DESI reference. We recognize that intrinsic variability of a compact narrow-line region may occur over the 12.2-year rest-frame interval between the SDSS and DESI observations. However, no alternative narrow emission line with adequate signal-to-noise is available as an independent calibration reference, while a direct comparison of the pre-normalization fluxes is affected by the different fiber apertures and absolute spectrophotometric calibrations of SDSS and DESI. We therefore retain [\ion{O}{3}]~$\lambda5007$ as a practical internal normalization reference. As demonstrated in the inset of Figure~\ref{fig:2}(a), the narrow [\ion{O}{3}] profiles are aligned after this normalization; by construction, the normalized spectra cannot independently test intrinsic [\ion{O}{3}] variability, and a narrow-line-region response cannot be excluded. The photometric baseline in Figure~\ref{fig:2}(b) combines archival SDSS and Pan-STARRS1 $g$- and $r$-band measurements with the later ZTF and \textit{NEOWISE}\citep{SDSSdr7,Chambers2016,NEOWISE2020,ZTF2025} monitoring, extending the coverage between the historical SDSS high state and the later low states \citep{SDSSdr7,Chambers2016,NEOWISE2020,ZTF2025}.

The automated spectral fitting tool \texttt{PyQSOFit} \citep{Pyqsofit1,Pyqsofit2} was implemented to decompose the DESI reference spectrum into the AGN and host-galaxy components. The host-galaxy spectrum extracted from this decomposition was subsequently modeled with the \texttt{STARLIGHT} code \citep{Starlight} using a combination of the \citet{BC03} evolutionary stellar templates under a Chabrier initial mass function \citep{Chab}. Given the moderate signal-to-noise ratio ($\text{S/N} \approx 11.45$) of the extracted stellar spectrum, we conservatively assigned a fixed systematic uncertainty of $0.1\,\text{dex}$ \citep{Starlight,Mateus2006} to the final host stellar mass estimation to account for potential template mismatches or base choices. This analysis yields a total host stellar mass of $\log (M_{\star}/M_{\odot}) = 11.12 \pm 0.10$. The same DESI-derived host component was adopted for the other spectroscopic epochs rather than fitting an independent host spectrum at each epoch.

\begin{figure*}[htbp]
\centering
\includegraphics[width=0.98\textwidth]{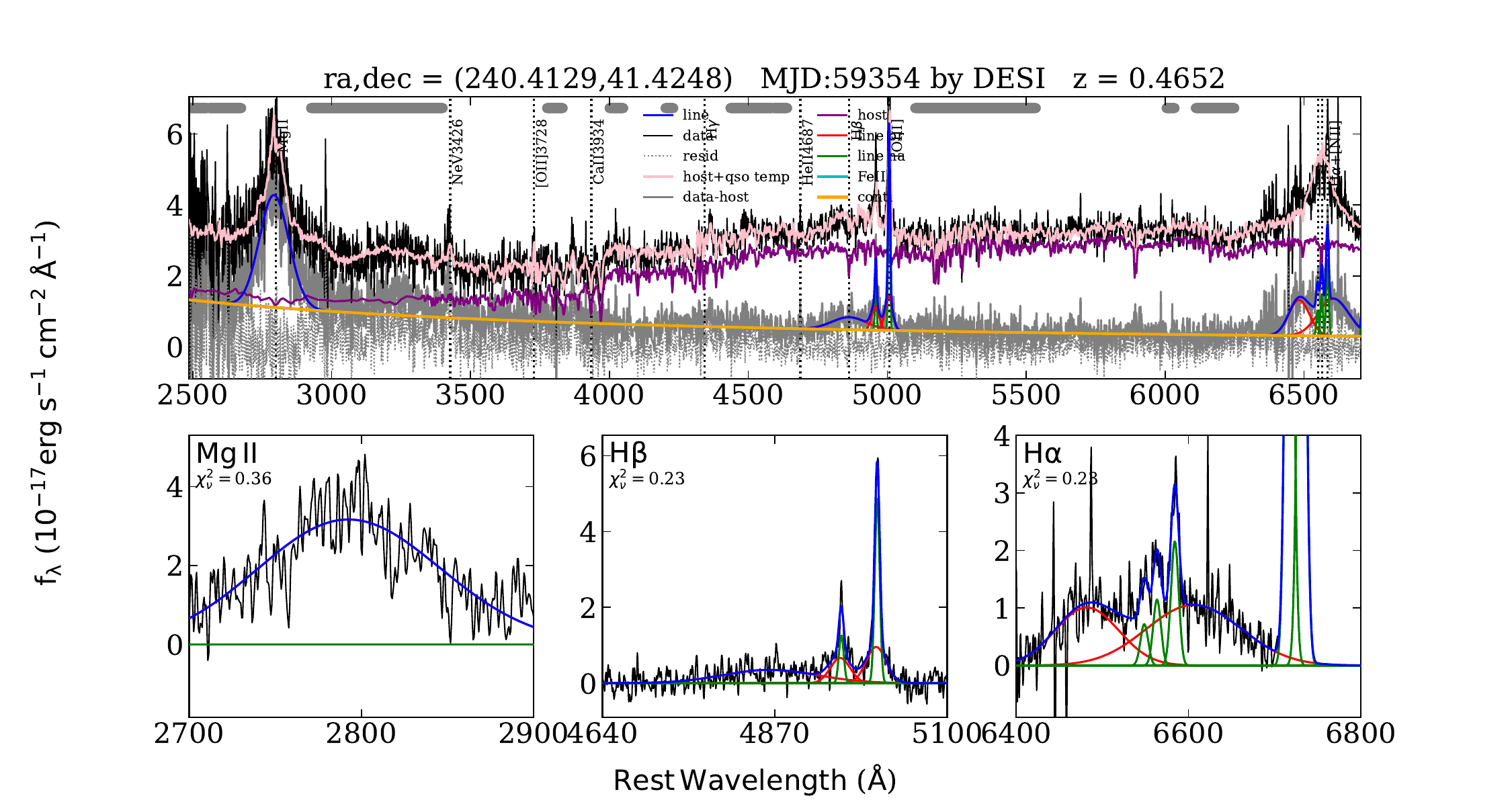}
\caption{Spectral decomposition and multi-component profile fitting of the baseline DESI spectrum (MJD 59354; $z = 0.4652$) performed via \texttt{PyQSOFit}. The primary panel illustrates the global rest-frame fit spanning the optical-to-UV range, showcasing the separation of the observed data into the power-law continuum, host galaxy starlight, and blended \ion{Fe}{2} pseudo-continuum. The surrounding sub-panels provide detailed close-up breakdowns of the primary emission line complexes: (a) the \ion{Mg}{2}~$\lambda2800$ region, (b) the \text{H}$\beta$-[\ion{O}{3}]~$\lambda\lambda4959,5007$ complex, and (c) the \text{H}$\alpha$-[\ion{N}{2}]-[\ion{S}{2}] complex (explicitly highlighted in the zoomed inset). In each sub-panel, the black solid line represents the host-subtracted AGN spectrum, the blue curve denotes the total model fit, and the green and red curves delineate individual narrow and broad emission line components, respectively.}
\label{fig:3}
\end{figure*}

\begin{figure*}[htbp]
\centering
\includegraphics[width=0.98\textwidth]{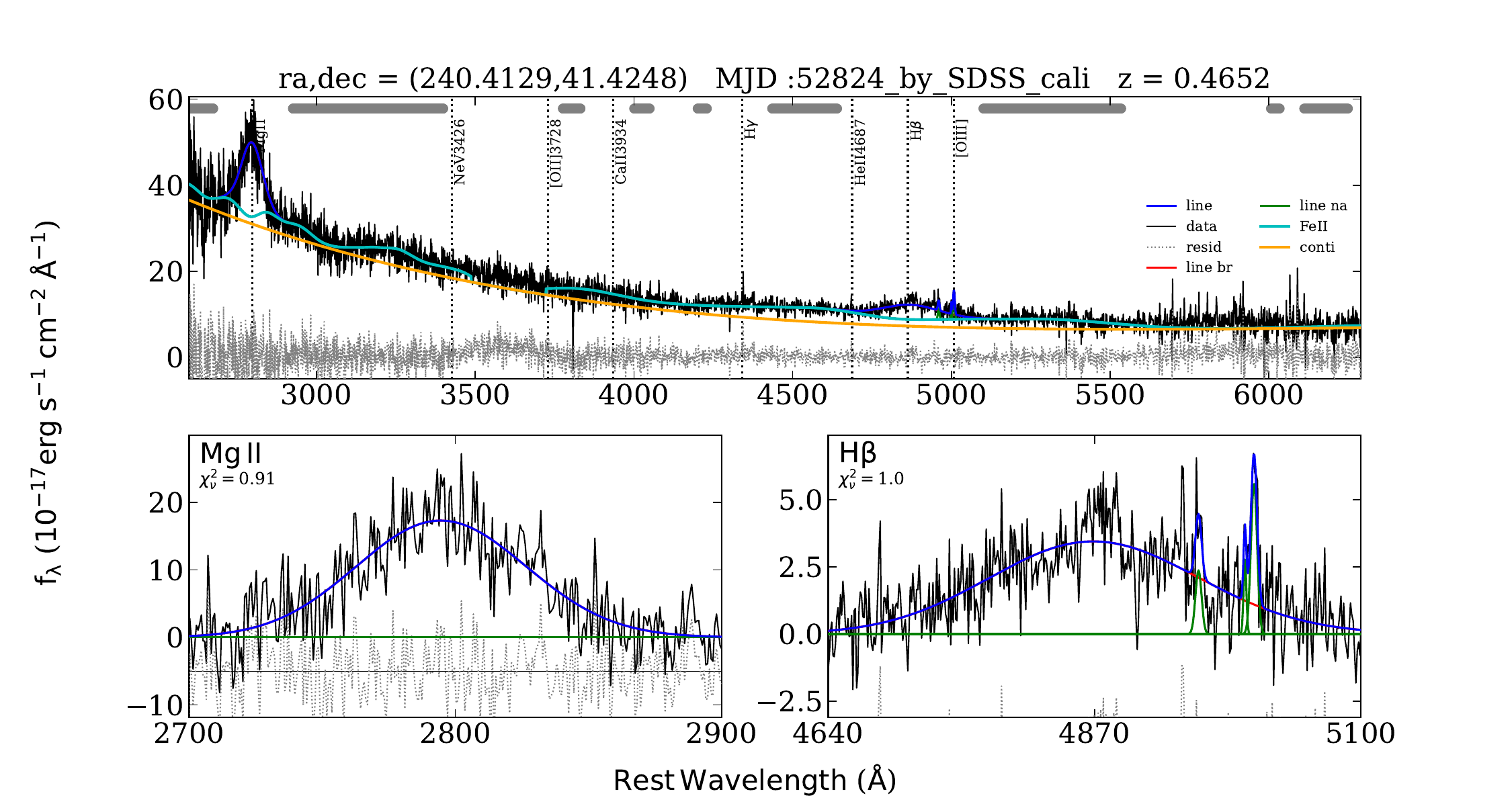}
\caption{Spectral decomposition and multi-component profile fitting of the SDSS no-host spectrum (MJD 52824) performed via \texttt{PyQSOFit}. The panels follow the same configuration and color schemes as described in Figure~\ref{fig:3}, capturing the target in its historical high state, where the broad Balmer emission lines and power-law continuum are stronger than in the later spectra.}
\label{fig:4}
\end{figure*}

Figures~\ref{fig:3} and \ref{fig:4} present the detailed emission-line profile decompositions for the DESI baseline spectrum and the historical SDSS high-state spectrum, respectively. Utilizing these recipes for our high-state spectrum, the broad emission lines exhibit large velocity widths, with $\text{FWHM}_{\text{H}\beta} = 13264.42 \pm 709.59\,\text{km~s}^{-1}$ and $\text{FWHM}_{\text{Mg~II}} = 8125.08 \pm 338.17\,\text{km~s}^{-1}$. These parameters yield virial mass estimates of $\log (M_{\text{BH, H}\beta}/M_{\odot}) = 9.41 \pm 0.05$ and $\log (M_{\text{BH, Mg~II}}/M_{\odot}) = 9.05 \pm 0.04$, respectively. In the subsequent low state (DESI; MJD~59354), the broad $\text{H}\beta$ component is not significantly detected, whereas the broad \ion{Mg}{2} remains detectable with a measured velocity width of $\text{FWHM}_{\text{Mg~II}} = 13393.11 \pm 689.00\,\text{km~s}^{-1}$. The two emission lines therefore show different changes between the observed epochs.

\section{Data Analysis and Results} \label{sec:results}

\subsection{The LoBAL Incidence among CLAGNs} \label{subsec:statistical_implication}

To understand the macro-physical properties governing the global changing-look population, we contextualize our search results within the statistical framework of the underlying parent sample. Our systematic search for broad absorption line (BAL) signatures across a well-defined sample of 370 optically selected changing-look active galactic nuclei (CLAGNs) yields one LoBAL source, SDSS J160139.10+412529.3 (see Section~\ref{sec:data}). Using the Wilson score interval \citep{Wilson1927}, this detection corresponds to an observed LoBAL incidence of $f_{\rm LoBAL}=0.27\%_{-0.17\%}^{+0.44\%}$ within the CLAGN sample.

To examine whether the observed LoBAL incidence is related to the physical properties of the changing-look population, we use the quasar properties provided by \citet{Guo2025}. During their historical bright states, the CLAGN sample has a mean continuum luminosity of $\langle \log (\lambda L_{\lambda,5100}/\text{erg~s}^{-1}) \rangle = 44.249 \pm 0.019$, a mean virial black hole mass of $\langle \log (M_{\rm BH}/M_{\odot}) \rangle = 8.535 \pm 0.025$, and a mean Eddington ratio of $\langle \log \lambda_{\rm Edd} \rangle = -1.387 \pm 0.019$.

\begin{figure*}[htbp]
\centering
\includegraphics[width=0.98\textwidth]{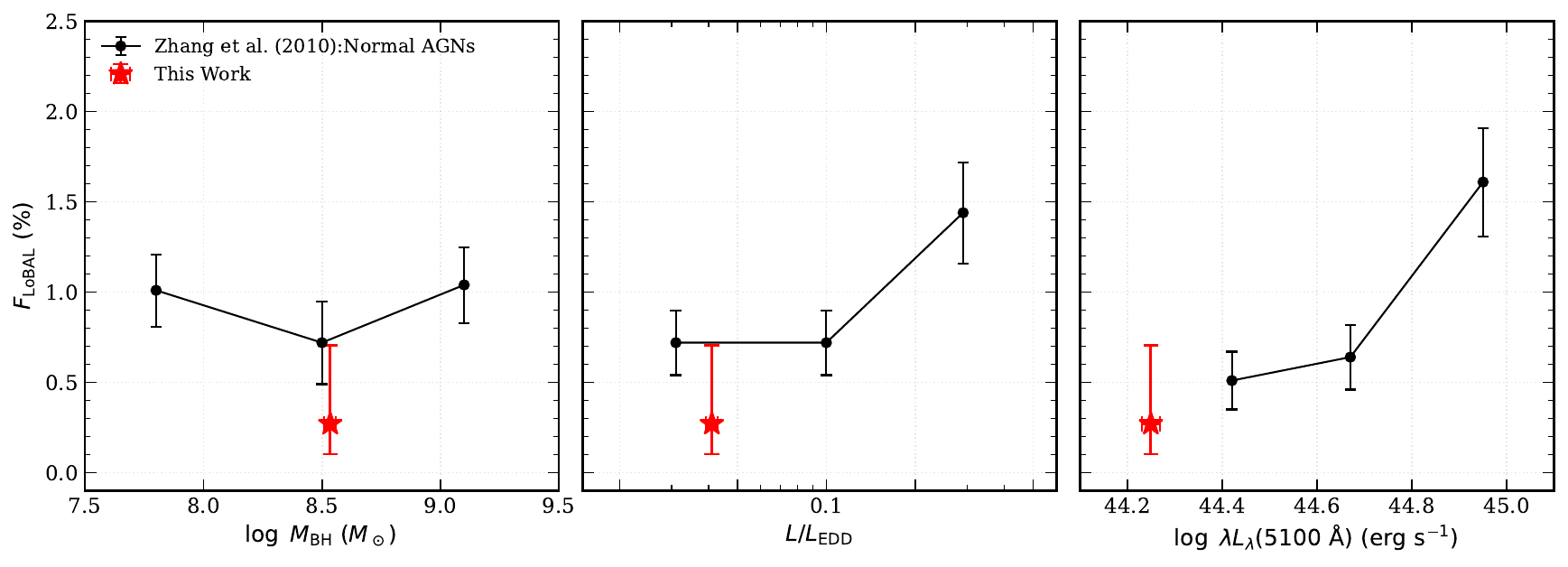}
\caption{The LoBAL fraction ($f_{\rm LoBAL}$) as a function of fundamental black hole accretion parameters. From left to right, the panels display the LoBAL fraction versus virial black hole mass ($\log M_{\rm BH}$), Eddington ratio ($\lambda_{\rm Edd} \equiv L_{\rm bol}/L_{\rm Edd}$), and rest-frame continuum luminosity ($\log \lambda L_{\lambda}$ at $5100\,\mathrm{\AA}$). The solid black circles connected by solid lines represent the uniform low-redshift empirical baselines derived from the standard LoBAL quasar sample of \citet{Zhang2010}. The red star marks the statistical constraint determined from this work based on one detection among 370 CLAGNs, with the asymmetric vertical error bars calculated using the Wilson score interval. The CLAGN measurement has a lower central value than the reference quasar baselines but remains statistically consistent with the empirical trends within approximately $1.5\sigma$.}
\label{fig:5}
\end{figure*}

We directly compare our result with the uniform low-redshift LoBAL statistical baseline established by \citet{Zhang2010}. As illustrated in Figure~\ref{fig:5}, we compare the CLAGN constraint with the empirical LoBAL fractions as functions of $M_{\rm BH}$, $\lambda_{\rm Edd}$, and $\lambda L_{\lambda}(5100\,\text{\AA})$. \citet{Zhang2010} found that the LoBAL fraction increases toward higher Eddington ratios and higher optical luminosities, while showing no statistically significant dependence on black hole mass.

The observed CLAGN LoBAL fraction of $0.27\%$ lies below the central values of the \citet{Zhang2010} LoBAL quasar distributions in all three parameter spaces (Figure~\ref{fig:5}). Because the LoBAL fraction decreases toward lower luminosities and lower Eddington ratios \citep{Zhang2010}, the low incidence measured in the CLAGN sample may partly reflect its generally low accretion rates. The CLAGN measurement follows the direction of the empirical trends found for ordinary LoBAL quasars. Although its central value is lower, it remains statistically consistent with the quasar baseline within approximately $1.5\sigma$. The difference is therefore not statistically significant.

The measured fraction represents the incidence of detectable LoBAL features within the available multi-epoch spectra and may be affected by observational selection. The requirement that the \ion{Mg}{2} region be covered in both SDSS/BOSS and DESI restricts the analysis to $z\geq0.33$, while the sensitivity to weak absorption varies with spectral signal-to-noise ratio. The available wavelength coverage does not provide uniform access to higher-ionization diagnostic lines such as \ion{C}{4} or \ion{Si}{4}. In addition, weak or transient BAL features may not be detected when they occur between the available spectroscopic epochs. These effects could cause the observed LoBAL fraction to underestimate the intrinsic incidence of low-ionization BAL outflows among CLAGNs.

The statistical consistency with the ordinary-quasar baseline indicates that the current measurement does not require an additional orientation effect beyond the dependence on accretion properties. The lower central value may still allow a contribution from viewing geometry, but the present $\sim1.5\sigma$ difference is insufficient to distinguish such a contribution from statistical uncertainty and observational selection effects. A larger CLAGN sample with more uniform spectral coverage and sensitivity will be required to determine whether viewing geometry further modifies the LoBAL incidence.

\subsection{Spectral Decomposition and Physical Parameters} \label{subsec:spectral_param}

Using the multi-epoch spectral processing and decomposition described in Section~\ref{app:spectral_details}, we analyzed the SDSS (MJD~52824), DESI (MJD~59354), and SDSS-V/BOSS (MJD~60114) spectra of J160139.10+412529.3. Briefly, the host-galaxy component extracted from the DESI reference spectrum was modeled with \texttt{STARLIGHT} and then adopted for the other spectroscopic epochs. The DESI-derived host spectrum yields a stellar mass of $\log(M_{\star}/M_{\odot})=11.12\pm0.10$. The resulting spectral decompositions are presented in Figures~\ref{fig:3} and \ref{fig:4}; here we focus on the derived black hole and accretion parameters.

Based on the Single-Epoch (SE) virial black hole mass ($M_{\rm BH}$) framework parameterized by \citet{Shen2011}, the physical properties derived from the high-state SDSS spectrum are $\log(L_{5100}/\text{erg~s}^{-1})=44.50\pm0.04$ and $\log(L_{3000}/\text{erg~s}^{-1})=44.80\pm0.01$. The broad emission-line profiles have widths of $\text{FWHM}_{\text{H}\beta}\sim13,260\,\text{km~s}^{-1}$ and $\text{FWHM}_{\text{Mg~II}}\sim8,130\,\text{km~s}^{-1}$, yielding virial mass estimates of $\log(M_{\rm BH,H\beta}/M_{\odot})=9.41\pm0.05$ and $\log(M_{\rm BH,Mg~II}/M_{\odot})=9.05\pm0.04$, respectively (see Table~\ref{tab:properties}).

We adopt the black hole masses measured from the high-state SDSS spectrum as fixed reference values when calculating the Eddington ratios at all epochs. We do not derive independent virial black hole masses from the lower-state spectra because single-epoch mass estimates may be unreliable when the continuum and broad emission-line profiles have changed substantially. The blank entries in the low-state black hole mass columns of Table~\ref{tab:properties} therefore indicate that no independent low-state mass was measured.

Adopting a bolometric correction of $L_{\rm bol}=9.26\times L_{5100}$ \citep{Richards2006}, and using the high-state H$\beta$-based black hole mass as the fixed mass scale, the Eddington ratio decreases from $\log\lambda_{\rm Edd,H\beta}=-2.05\pm0.07$ in the high state to $-3.29\pm0.05$ in the lower state. Similarly, using the high-state \ion{Mg}{2}-based black hole mass for both epochs, the inferred Eddington ratio decreases from $\log\lambda_{\rm Edd,Mg~II}=-1.70\pm0.05$ to $-2.93\pm0.04$. These changes, together with the approximately $2$~mag decline in the optical light curve, are consistent with a substantial change in the accretion state.

\begin{table*}[t!]
\centering
\caption{Spectral Measurements and Derived Physical Properties}
\label{tab:properties}
\footnotesize
\setlength{\tabcolsep}{4pt}
\begin{tabular}{l ccccccc}
\toprule
State & $\log L_{5100}$ & $\text{FWHM}_{\text{H}\beta}$ & $\text{FWHM}_{\text{Mg~II}}$ & $\log M_{\text{BH, H}\beta}$ & $\log M_{\text{BH, Mg~II}}$ & $\log \lambda_{\text{Edd, H}\beta}$ & $\log \lambda_{\text{Edd, Mg~II}}$ \\
 & ($\text{erg~s}^{-1}$) & ($\text{km~s}^{-1}$) & ($\text{km~s}^{-1}$) & ($\log M_{\odot}$) & ($\log M_{\odot}$) & & \\
\midrule
High & \textbf{44.50} $\pm$ \textbf{0.04} & \textbf{13264.42} $\pm$ \textbf{709.59} & \textbf{8125.08} $\pm$ \textbf{338.17} & \textbf{9.41} $\pm$ \textbf{0.05} & \textbf{9.05} $\pm$ \textbf{0.04} & \textbf{-2.05} $\pm$ \textbf{0.068} & \textbf{-1.70} $\pm$ \textbf{0.054} \\
Low  & \textbf{43.26} $\pm$ \textbf{0.01} & \textbf{-} & \textbf{13393.11} $\pm$ \textbf{689.00} & \textbf{-} & \textbf{-} & \textbf{-3.29} $\pm$ \textbf{0.052} & \textbf{-2.93} $\pm$ \textbf{0.038} \\
\bottomrule
\end{tabular}
\tablecomments{The black hole masses are derived from the high-state SDSS spectrum and are adopted as fixed reference values when calculating the Eddington ratios in both states. The dashes in the low-state black hole mass columns indicate that no independent low-state virial mass estimates are reported.}
\end{table*}

\section{Discussion } \label{sec:discus}

\subsection{Ion Identification and Absorption-Profile Modeling}

A primary challenge in interpreting the outflow profile is identifying the absorbing ion, because the available wavelength coverage does not include shorter-wavelength diagnostic lines that could distinguish between the \ion{Mg}{2} and \ion{Fe}{2} interpretations. As shown in Figure~\ref{fig:1}, the primary absorption feature spans approximately 15 pixels ($\approx10\,\text{\AA}$ in the rest frame). A second, shallower absorption feature, approximately four pixels wide and broader than the instrumental resolution, is present on the red side of the primary trough.

We first test whether the two absorption features can be represented by the \ion{Mg}{2}~$\lambda\lambda2796,2803$ doublet. The optical-depth profile is modeled as
\begin{equation}
\tau (\lambda) = \tau_{0}\exp\left[-\frac{(\lambda-\mu_{1})^{2}}{2\sigma_{\lambda}^{2}}\right] \\
+\frac{\tau_{0}}{2}\exp\left[-\frac{(\lambda-\mu_{2})^{2}}{2\sigma_{\lambda}^{2}}\right],
\end{equation}
where $\mu_i=\lambda_i(1-v/c)$. The two components share the same velocity shift $v$ and Gaussian width $\sigma_{\lambda}$, and their central optical depths are fixed at the theoretical ratio $\tau_{2796}:\tau_{2803}=2:1$, consistent with the ratio of their oscillator strengths \citep{Morton2003}. The model flux is
\begin{equation}
F_{\rm model}(\lambda)=F_{\rm cont}(\lambda)\exp[-\tau_{\rm Mg\,II}(\lambda)],
\end{equation}
where $F_{\rm cont}(\lambda)$ is the reconstructed unabsorbed continuum. The free absorption parameters are $v$, $\sigma_{\lambda}$, and the central optical depth $\tau_{0}$ of the $\lambda2796$ component. This constrained doublet model reproduces the shallow red feature and the deepest part of the primary trough, yielding $\tau_{0}=0.64$. The agreement with the expected doublet separation and optical-depth ratio supports an \ion{Mg}{2} interpretation but does not uniquely establish it.

We therefore also test a single-transition \ion{Fe}{2} interpretation for the primary trough, adopting the \ion{Fe}{2}~$\lambda2755.75$ transition. The normalized residual intensity is modeled with a partial-covering prescription,
\begin{equation}
\frac{F(v)}{F_{\rm cont}(v)}=1-C+C\exp[-\tau(v)],
\end{equation}
where the covering factor $C$ and optical-depth profile $\tau(v)$ are fitted over the velocity bins within the primary trough. We use the same fitting framework as for a doublet but set the secondary-to-primary line-strength ratio and doublet separation to zero, thereby reducing the model to a single absorption transition. No Gaussian form is imposed on $\tau(v)$. The fit yields a covering factor of $C=0.567$.

The covering-factor-weighted ionic column-density profile is calculated from the fitted optical depth using the standard relation \citep{Savage1991}
\begin{equation}
N_{\rm Fe\,II}(v)=\frac{3.768\times10^{14}}{f_{2755}\lambda_{2755}}C\tau(v)
\quad \text{cm}^{-2}\,(\text{km\,s}^{-1})^{-1},
\end{equation}
where $\lambda_{2755}=2755.75$~\text{\AA} and $f_{2755}=0.306$. The velocity coordinate is defined as
\begin{equation}
v=c\frac{\lambda-\lambda_{2755}}{\lambda_{2755}},
\end{equation}
and the total covering-factor-weighted column density is obtained by integrating over the fitted velocity interval,
\begin{equation}
N_{\rm Fe\,II}=\frac{3.768\times10^{14}}{f_{2755}\lambda_{2755}}\int C\tau(v)\,dv.
\end{equation}
This procedure gives $N_{\rm Fe\,II}=2.43\times10^{14}\,\text{cm}^{-2}$ for the absorption component with a blueshift magnitude of approximately $12,000\,\text{km\,s}^{-1}$.

The constrained \ion{Mg}{2} doublet model describes both the primary trough and the shallower red feature, whereas the \ion{Fe}{2} model provides a single-transition description of the primary trough. The available spectra do not provide sufficient independent transitions to distinguish uniquely between these interpretations. We therefore refer to the observed feature as an \ion{Mg}{2}/\ion{Fe}{2} BAL rather than assigning it exclusively to \ion{Mg}{2}. This ionic ambiguity does not affect the classification of the source as a LoBAL: an \ion{Mg}{2} origin corresponds to a LoBAL, while a dominant \ion{Fe}{2} contribution would place the source in the FeLoBAL subclass. However, the available wavelength coverage cannot determine whether high-ionization BALs are also present. Ultraviolet spectroscopy covering transitions such as \ion{C}{4}, \ion{Si}{4}, and \ion{Al}{3} is required to constrain the full ionization structure of the outflow.

\subsection{The Origins of Emission-Line and LoBAL Variability}

The high-state SDSS spectrum shows both broad emission lines and a low-ionization BAL feature, establishing that the two phenomena were present in the same source during that epoch. Across the subsequent spectroscopic epochs, the broad H$\beta$ emission and the \ion{Mg}{2}/\ion{Fe}{2} BAL feature both become substantially weaker. The source can therefore be regarded as a LoBAL--CLAGN coexistence case in its high state, while its multi-epoch evolution also represents BAL variability accompanying a changing-look transition. These descriptions refer to different aspects of the same source and are not mutually exclusive.

The broad-emission-line transition is accompanied by a long-term optical continuum decline of approximately $2$~mag from the historical SDSS high state to the later low states (Figure~\ref{fig:2}). Broad H$\beta$ becomes undetectable in the later spectra, while a broad H$\alpha$ component remains visible in the DESI epoch. Together with the multi-band continuum decline, this behavior favors an intrinsic change in the ionizing continuum associated with a change in the accretion state, rather than a simple line-of-sight obscuration scenario \citep{MacLeod2016,MacLeod2019}.

The observed line-dependent evolution, in which broad H$\beta$ becomes undetectable while broad \ion{Mg}{2} remains present, is consistent with the predictions of the Locally Optimally Emitting Cloud (LOC) photoionization models \citep{GuoCloudy2020}. In these models, H$\beta$ and \ion{Mg}{2} have different responsivities to a declining ionizing continuum because they are preferentially emitted under different physical conditions within the BLR. Consequently, H$\beta$ can weaken below the detection threshold while \ion{Mg}{2} remains detectable.

Recent wavelength-resolved reverberation-mapping studies provide additional evidence that low-ionization emission lines do not necessarily trace identical regions of the BLR. \citet{Prince2023} found that the relative locations of the \ion{Mg}{2} and UV \ion{Fe}{2} emission regions can differ between sources, although some inferred differences are not statistically significant. \citet{Zajacek2024} further found that the \ion{Mg}{2} radius--luminosity relation is flatter than those of H$\beta$ and the optical and UV \ion{Fe}{2} emission. These results provide broader context for line-dependent BLR structure, while the specific H$\beta$/\ion{Mg}{2} evolution observed here is interpreted primarily through the LOC model.

The substantial weakening of the \ion{Mg}{2}/\ion{Fe}{2} BAL feature may result from either transverse motion of the absorbing gas or a change in its ionization state. In the transverse-motion scenario, an absorbing structure moves partially or completely out of the line of sight between the observed epochs. Given the approximately one-decade rest-frame interval between the high and low states, such motion could reduce the line-of-sight covering factor. This interpretation does not require the BAL variation and the changing-look transition to share the same physical cause.

Alternatively, the BAL variation may be related to photoionization changes driven by the declining ionizing continuum. A change in the ionizing flux can alter the ionic populations and optical depths of the absorbing gas, causing a BAL trough to strengthen or weaken depending on the initial ionization state and gas density \citep{Wang2015,He2017,He2025}. Under this interpretation, the temporal association between the continuum decline and BAL weakening may reflect a response of the outflowing gas to the changing central source\citep{he2024}. However, the present spectra do not uniquely determine the ionization state of the absorber or predict the sign of its response without additional ultraviolet ionic diagnostics.

The current observations therefore do not distinguish between transverse motion and photoionization variability as the origin of the BAL change. Moreover, the temporal association between the emission-line transition and BAL weakening does not by itself demonstrate either a common physical origin or complete physical decoupling. Additional ultraviolet spectroscopy and more densely sampled spectroscopic monitoring will be required to determine the ionic composition of the absorber and its response to changes in the ionizing continuum.

\begin{acknowledgments}
We are grateful to Tek Prasad Adhikari for stimulating discussions. Z. C. He is supported by the National Natural Science Foundation of China (Grant Nos. 12633004, 12192220, 12192221) and the National SKA Program of China (No. 2025SKA0130100).
\end{acknowledgments}

\clearpage

\bibliography{ref}{}

@article{Mereghetti:2021aa,
	adsurl = {https://ui.adsabs.harvard.edu/abs/2021ExA...tmp..143M},
	archiveprefix = {arXiv},
	author = {{Mereghetti}, S. and {Balman}, S. and {Caballero-Garcia}, M. and {Del Santo}, M. and {Doroshenko}, V. and {Erkut}, M.~H. and {Hanlon}, L. and {Hoeflich}, P. and {Markowitz}, A. and {Osborne}, J.~P. and {Pian}, E. and {Rivera Sandoval}, L. and {Webb}, N. and {Amati}, L. and {Ambrosi}, E. and {Beardmore}, A.~P. and {Blain}, A. and {Bozzo}, E. and {Burderi}, L. and {Campana}, S. and {Casella}, P. and {D'A{\'\i}}, A. and {D'Ammando}, F. and {De Colle}, F. and {Della Valle}, M. and {De Martino}, D. and {Di Salvo}, T. and {Doyle}, M. and {Esposito}, P. and {Frontera}, F. and {Gandhi}, P. and {Ghisellini}, G. and {Gotz}, D. and {Grinberg}, V. and {Guidorzi}, C. and {Hudec}, R. and {Iaria}, R. and {Izzo}, L. and {Jaisawal}, G.~K. and {Jonker}, P.~G. and {Kong}, A.~K.~H. and {Krumpe}, M. and {Kumar}, P. and {Manousakis}, A. and {Marino}, A. and {Martin-Carrillo}, A. and {Mignani}, R. and {Miniutti}, G. and {Mundell}, C.~G. and {Mukai}, K. and {Nucita}, A.~A. and {O'Brien}, P.~T. and {Orlandini}, M. and {Orio}, M. and {Palazzi}, E. and {Papitto}, A. and {Pintore}, F. and {Piranomonte}, S. and {Porquet}, D. and {Ricci}, C. and {Riggio}, A. and {Rigoselli}, M. and {Rodriguez}, J. and {Saha}, T. and {Sanna}, A. and {Santangelo}, A. and {Saxton}, R. and {Sidoli}, L. and {Stiele}, H. and {Tagliaferri}, G. and {Tavecchio}, F. and {Tiengo}, A. and {Tsygankov}, S. and {Turriziani}, S. and {Wijnands}, R. and {Zane}, S. and {Zhang}, B.},
	doi = {10.1007/s10686-021-09809-6},
	eprint = {2104.09533},
	journal = {Experimental Astronomy},
	month = nov,
	primaryclass = {astro-ph.IM},
	title = {{Time domain astronomy with the THESEUS satellite}},
	year = 2021}

@article{Graham:2020nz,
	adsurl = {https://ui.adsabs.harvard.edu/abs/2020MNRAS.491.4925G},
	archiveprefix = {arXiv},
	author = {{Graham}, Matthew J. and {Ross}, Nicholas P. and {Stern}, Daniel and {Drake}, Andrew J. and {McKernan}, Barry and {Ford}, K.~E. Saavik and {Djorgovski}, S.~G. and {Mahabal}, Ashish A. and {Glikman}, Eilat and {Larson}, Steve and {Christensen}, Eric},
	doi = {10.1093/mnras/stz3244},
	eprint = {1905.02262},
	journal = {\mnras},
	month = feb,
	number = {4},
	pages = {4925-4948},
	primaryclass = {astro-ph.GA},
	title = {{Understanding extreme quasar optical variability with CRTS - II. Changing-state quasars}},
	volume = {491},
	year = 2020}

@article{Ricci2023,
	adsurl = {https://ui.adsabs.harvard.edu/abs/2023NatAs...7.1282R},
	archiveprefix = {arXiv},
	author = {{Ricci}, Claudio and {Trakhtenbrot}, Benny},
	doi = {10.1038/s41550-023-02108-4},
	eprint = {2211.05132},
	journal = {Nature Astronomy},
	month = nov,
	pages = {1282-1294},
	primaryclass = {astro-ph.GA},
	title = {{Changing-look active galactic nuclei}},
	volume = {7},
	year = 2023}

@article{MacLeod2016,
	adsurl = {https://ui.adsabs.harvard.edu/abs/2016MNRAS.457..389M},
	archiveprefix = {arXiv},
	author = {{MacLeod}, Chelsea L. and {Ross}, Nicholas P. and {Lawrence}, Andy and {Goad}, Mike and {Horne}, Keith and {Burgett}, William and {Chambers}, Ken C. and {Flewelling}, Heather and {Hodapp}, Klaus and {Kaiser}, Nick and {Magnier}, Eugene and {Wainscoat}, Richard and {Waters}, Christopher},
	doi = {10.1093/mnras/stv2997},
	eprint = {1509.08393},
	journal = {\mnras},
	month = mar,
	number = {1},
	pages = {389-404},
	primaryclass = {astro-ph.GA},
	title = {{A systematic search for changing-look quasars in SDSS}},
	volume = {457},
	year = 2016}

@article{MacLeod2019,
	adsurl = {https://ui.adsabs.harvard.edu/abs/2019ApJ...874....8M},
	archiveprefix = {arXiv},
	author = {{MacLeod}, Chelsea L. and {Green}, Paul J. and {Anderson}, Scott F. and {Bruce}, Alastair and {Eracleous}, Michael and {Graham}, Matthew and {Homan}, David and {Lawrence}, Andy and {LeBleu}, Amy and {Ross}, Nicholas P. and {Ruan}, John J. and {Runnoe}, Jessie and {Stern}, Daniel and {Burgett}, William and {Chambers}, Kenneth C. and {Kaiser}, Nick and {Magnier}, Eugene and {Metcalfe}, Nigel},
	doi = {10.3847/1538-4357/ab05e2},
	eid = {8},
	eprint = {1810.00087},
	journal = {\apj},
	month = mar,
	number = {1},
	pages = {8},
	primaryclass = {astro-ph.GA},
	title = {{Changing-look Quasar Candidates: First Results from Follow-up Spectroscopy of Highly Optically Variable Quasars}},
	volume = {874},
	year = 2019}

@article{guoHigh2020,
  title = {High-Redshift {{Extreme Variability Quasars}} from {{Sloan Digital Sky Survey Multiepoch Spectroscopy}}},
  author = {Guo, Hengxiao and Peng, Jiacheng and Zhang, Kaiwen and Burke, Colin J. and Liu, Xin and Sun, Mouyuan and Wang, Shu and Kong, Minzhi and Sheng, Zhenfeng and Wang, Tinggui and He, Zhicheng and Gu, Minfeng},
  year = 2020,
  month = dec,
  journal = {The Astrophysical Journal},
  volume = {905},
  number = {1},
  pages = {52},
  publisher = {The American Astronomical Society},
  issn = {0004-637X},
  doi = {10.3847/1538-4357/abc2ce},
  urldate = {2026-03-25},
  langid = {english}
}

@misc{guo2024,
  title = {The First Identification of {{Lyman}} \${$\alpha\$$} {{Changing-look Quasars}} at High-Redshift in {{DESI}}},
  author = {Guo, Wei-Jian and Pan, Zhiwei and Siudek, Ma{\l}gorzata and Aguilar, Jessica Nicole and Ahlen, Steven and Bianchi, Davide and Brooks, David and Claybaugh, Todd and Dawson, Kyle and de la Macorra, Axel and Doel, Peter and Fanning, Kevin and {Forero-Romero}, Jaime E. and Gazta{\~n}aga, Enrique and Gontcho, Satya Gontcho A. and Honscheid, Klaus and Kehoe, Robert and Kisner, Theodore and Lambert, Andrew and Landriau, Martin and Guillou, Laurent Le and Manera, Marc and Meisner, Aaron and Moustakas, John and {Mu{\~n}oz-Guti{\'e}rrez}, Andrea and Myers, Adam and Nie, Jundan and {Palanque-Delabrouille}, Nathalie and Poppett, Claire and Prada, Francisco and Rezaie, Mehdi and Rossi, Graziano and Sanchez, Eusebio and Schubnelll, Michael and Seo, Hee-Jong and Silber, Joseph Harry and Sprayberry, David and Tarl{\'e}, Gregory and Weaver, Benjamin Alan and Zhou, Zhimin and Zou, Hu},
  year = 2024,
  month = nov,
  number = {arXiv:2411.01949},
  eprint = {2411.01949},
  primaryclass = {astro-ph},
  publisher = {arXiv},
  doi = {10.48550/arXiv.2411.01949},
  urldate = {2025-11-17},
  archiveprefix = {arXiv}
}

@article{sheng2017,
  title = {Mid-Infrared {{Variability}} of {{Changing-look AGNs}}},
  author = {Sheng, Zhenfeng and Wang, Tinggui and Jiang, Ning and Yang, Chenwei and Yan, Lin and Dou, Liming and Peng, Bo},
  year = 2017,
  month = aug,
  journal = {The Astrophysical Journal Letters},
  volume = {846},
  number = {1},
  pages = {L7},
  publisher = {The American Astronomical Society},
  issn = {2041-8205},
  doi = {10.3847/2041-8213/aa85de},
  url = {https://doi.org/10.3847/2041-8213/aa85de},
  urldate = {2026-03-25},
  langid = {english}
}

@article{Kokubo:2020kr,
	adsurl = {https://ui.adsabs.harvard.edu/abs/2020MNRAS.491.4615K},
	archiveprefix = {arXiv},
	author = {{Kokubo}, Mitsuru and {Minezaki}, Takeo},
	doi = {10.1093/mnras/stz3397},
	eprint = {1904.08946},
	journal = {\mnras},
	month = feb,
	number = {4},
	pages = {4615-4633},
	primaryclass = {astro-ph.GA},
	title = {{Rapid luminosity decline and subsequent reformation of the innermost dust distribution in the changing-look AGN Mrk 590}},
	volume = {491},
	year = 2020}

@article{Burtscher16,
	adsurl = {http://adsabs.harvard.edu/abs/2016A%26A...586A..28B},
	archiveprefix = {arXiv},
	author = {{Burtscher}, L. and {Davies}, R.~I. and {Graci{\'a}-Carpio}, J. and {Koss}, M.~J. and {Lin}, M.-Y. and {Lutz}, D. and {Nandra}, P. and {Netzer}, H. and {Orban de Xivry}, G. and {Ricci}, C. and {Rosario}, D.~J. and {Veilleux}, S. and {Contursi}, A. and {Genzel}, R. and {Schnorr-M{\"u}ller}, A. and {Sternberg}, A. and {Sturm}, E. and {Tacconi}, L.~J.},
	doi = {10.1051/0004-6361/201527575},
	eid = {A28},
	eprint = {1511.05566},
	journal = {\aap},
	month = feb,
	pages = {A28},
	title = {{On the relation of optical obscuration and X-ray absorption in Seyfert galaxies}},
	volume = 586,
	year = 2016}

@article{Guo2025,
  title = {Changing-Look {{Active Galactic Nuclei}} from the {{Dark Energy Spectroscopic Instrument}}. {{II}}. {{Statistical Properties}} from the {{First Data Release}}},
  author = {Guo, Wei-Jian and Zou, Hu and Greenwell, Claire L. and Alexander, David M. and Fawcett, Victoria A. and Pan, Zhiwei and Siudek, Ma{\l}gorzata and Aguilar, Jessica Nicole and Ahlen, Steven and Brooks, David and Claybaugh, Todd and Dawson, Kyle and {de la Macorra}, Axel and Doel, Peter and {Font-Ribera}, Andreu and Gazta{\~n}aga, Enrique and Gontcho A Gontcho, Satya and Gutierrez, Gaston and Kehoe, Robert and Kisner, Theodore and Landriau, Martin and Le Guillou, Laurent and Manera, Marc and Meisner, Aaron and Miquel, Ramon and Moustakas, John and Prada, Francisco and Rossi, Graziano and Sanchez, Eusebio and Schubnell, Michael and Sprayberry, David and Sui, Jipeng and Tarl{\'e}, Gregory and Weaver, Benjamin Alan and Xiao, Yun-Ao and Zou, Siwei},
  year = 2025,
  month = may,
  journal = {The Astrophysical Journal Supplement Series},
  volume = {278},
  number = {1},
  pages = {28},
  publisher = {The American Astronomical Society},
  issn = {0067-0049},
  doi = {10.3847/1538-4365/adc124},
  url = {https://doi.org/10.3847/1538-4365/adc124},
  urldate = {2025-10-14},
  langid = {english}
}

@article{GuoCloudy2020,
  title = {Understanding {{Broad Mg}} Ii {{Variability}} in {{Quasars}} with {{Photoionization}}: {{Implications}} for {{Reverberation Mapping}} and {{Changing-look Quasars}}},
  shorttitle = {Understanding {{Broad Mg}} Ii {{Variability}} in {{Quasars}} with {{Photoionization}}},
  author = {Guo, Hengxiao and Shen, Yue and He, Zhicheng and Wang, Tinggui and Liu, Xin and Wang, Shu and Sun, Mouyuan and Yang, Qian and Kong, Minzhi and Sheng, Zhenfeng},
  year = 2020,
  month = jan,
  journal = {The Astrophysical Journal},
  volume = {888},
  number = {2},
  pages = {58},
  publisher = {The American Astronomical Society},
  issn = {0004-637X},
  doi = {10.3847/1538-4357/ab5db0},
  url = {https://doi.org/10.3847/1538-4357/ab5db0},
  urldate = {2025-12-26},
  langid = {english}
}

@article{Vivek2025,
  title = {Broad Absorption Line Outflows Are Common in Low {{Eddington}} Ratio Active Galactic Nuclei},
  author = {Vivek, M. and Wylezalek, Dominika},
  year = 2025,
  month = mar,
  journal = {Astronomy \& Astrophysics},
  volume = {695},
  pages = {L22},
  issn = {0004-6361, 1432-0746},
  doi = {10.1051/0004-6361/202453352},
  url = {https://www.aanda.org/10.1051/0004-6361/202453352},
  urldate = {2025-04-09},
  copyright = {https://creativecommons.org/licenses/by/4.0},
  langid = {english}
}

@ARTICLE{Starlight,
       author = {{Cid Fernandes}, Roberto and {Mateus}, Ab{\'\i}lio and {Sodr{\'e}}, Laerte and {Stasi{\'n}ska}, Gra{\.z}yna and {Gomes}, Jean M.},
        title = "{Semi-empirical analysis of Sloan Digital Sky Survey galaxies - I. Spectral synthesis method}",
      journal = {\mnras},
         year = 2005,
        month = apr,
       volume = {358},
       number = {2},
        pages = {363-378},
          doi = {10.1111/j.1365-2966.2005.08752.x},
archivePrefix = {arXiv},
       eprint = {astro-ph/0412481},
 primaryClass = {astro-ph},
       adsurl = {https://ui.adsabs.harvard.edu/abs/2005MNRAS.358..363C}
}

@ARTICLE{Mateus2006,
       author = {{Mateus}, Ab{\'\i}lio and {Sodr{\'e}}, Laerte and {Cid Fernandes}, Roberto and {Stasi{\'n}ska}, Gra{\.z}yna and {Schoenell}, William and {Gomes}, Jean M.},
        title = "{Semi-empirical analysis of Sloan Digital Sky Survey galaxies - II. The bimodality of the galaxy population revisited}",
      journal = {\mnras},
         year = 2006,
        month = aug,
       volume = {370},
       number = {2},
        pages = {721-737},
          doi = {10.1111/j.1365-2966.2006.10565.x},
archivePrefix = {arXiv},
       eprint = {astro-ph/0511578},
 primaryClass = {astro-ph},
       adsurl = {https://ui.adsabs.harvard.edu/abs/2006MNRAS.370..721M}
}

@ARTICLE{BC03,
       author = {{Bruzual}, G. and {Charlot}, S.},
        title = "{Stellar population synthesis at the resolution of 2003}",
      journal = {\mnras},
         year = 2003,
        month = oct,
       volume = {344},
       number = {4},
        pages = {1000-1028},
          doi = {10.1046/j.1365-8711.2003.06897.x},
archivePrefix = {arXiv},
       eprint = {astro-ph/0309134},
 primaryClass = {astro-ph},
       adsurl = {https://ui.adsabs.harvard.edu/abs/2003MNRAS.344.1000B}
}

@ARTICLE{Chab,
       author = {{Chabrier}, Gilles},
        title = "{Galactic Stellar and Substellar Initial Mass Function}",
      journal = {\pasp},
         year = 2003,
        month = jul,
       volume = {115},
       number = {809},
        pages = {763-795},
          doi = {10.1086/376392},
archivePrefix = {arXiv},
       eprint = {astro-ph/0304382},
 primaryClass = {astro-ph},
       adsurl = {https://ui.adsabs.harvard.edu/abs/2003PASP..115..763C}
}

@software{Pyqsofit1,
       author = {{Guo}, Hengxiao and {Shen}, Yue and {Wang}, Shu},
        title = "{PyQSOFit: Python code to fit the spectrum of quasars}",
 howpublished = {Astrophysics Source Code Library, record ascl:1809.008},
         year = 2018,
        month = sep,
          eid = {ascl:1809.008},
archivePrefix = {ascl},
       eprint = {1809.008},
       adsurl = {https://ui.adsabs.harvard.edu/abs/2018ascl.soft09008G}
}

@ARTICLE{Pyqsofit2,
       author = {{Shen}, Yue and {Hall}, Patrick B. and {Horne}, Keith and {Zhu}, Guangtun and {McGreer}, Ian and {Simm}, Torben and {Trump}, Jonathan R. and {Kinemuchi}, Karen and {Brandt}, W.~N. and {Green}, Paul J. and {Grier}, C.~J. and {Guo}, Hengxiao and {Ho}, Luis C. and {Homayouni}, Yasaman and {Jiang}, Linhua and {I-Hsiu Li}, Jennifer and {Morganson}, Eric and {Petitjean}, Patrick and {Richards}, Gordon T. and {Schneider}, Donald P. and {Starkey}, D.~A. and {Wang}, Shu and {Chambers}, Ken and {Kaiser}, Nick and {Kudritzki}, Rolf-Peter and {Magnier}, Eugene and {Waters}, Christopher},
        title = "{The Sloan Digital Sky Survey Reverberation Mapping Project: Sample Characterization}",
      journal = {\apjs},
         year = 2019,
        month = apr,
       volume = {241},
       number = {2},
          eid = {34},
        pages = {34},
          doi = {10.3847/1538-4365/ab074f},
archivePrefix = {arXiv},
       eprint = {1810.01447},
 primaryClass = {astro-ph.GA},
       adsurl = {https://ui.adsabs.harvard.edu/abs/2019ApJS..241...34S}
}

@article{Shen2011,
  title = {A {{Catalog}} of {{Quasar Properties}} from {{Sloan Digital Sky Survey Data Release}} 7},
  author = {Shen, Yue and Richards, Gordon T. and Strauss, Michael A. and Hall, Patrick B. and Schneider, Donald P. and Snedden, Stephanie and Bizyaev, Dmitry and Brewington, Howard and Malanushenko, Viktor and Malanushenko, Elena and Oravetz, Dan and Pan, Kaike and Simmons, Audrey},
  year = 2011,
  month = jun,
  journal = {The Astrophysical Journal Supplement Series},
  volume = {194},
  pages = {45},
  publisher = {IOP},
  issn = {0067-0049},
  doi = {10.1088/0067-0049/194/2/45},
  url = {https://ui.adsabs.harvard.edu/abs/2011ApJS..194...45S},
  urldate = {2026-04-02}
}

@ARTICLE{Richards2006,
       author = {{Richards}, Gordon T. and {Lacy}, Mark and {Storrie-Lombardi}, Lisa J. and {Hall}, Patrick B. and {Gallagher}, S.~C. and {Hines}, Dean C. and {Fan}, Xiaohui and {Papovich}, Casey and {Vanden Berk}, Daniel E. and {Trammell}, George B. and {Schneider}, Donald P. and {Vestergaard}, Marianne and {York}, Donald G. and {Jester}, Sebastian and {Anderson}, Scott F. and {Budav{\'a}ri}, Tam{\'a}s and {Szalay}, Alexander S.},
        title = "{Spectral Energy Distributions and Multiwavelength Selection of Type 1 Quasars}",
      journal = {\apjs},
         year = 2006,
        month = oct,
       volume = {166},
       number = {2},
        pages = {470-497},
          doi = {10.1086/506525},
archivePrefix = {arXiv},
       eprint = {astro-ph/0601558},
 primaryClass = {astro-ph},
       adsurl = {https://ui.adsabs.harvard.edu/abs/2006ApJS..166..470R}
}

@misc{Zhu2016,
  title = {Nonnegative {{Matrix Factorization}} ({{NMF}}) with {{Heteroscedastic Uncertainties}} and {{Missing}} Data},
  author = {Zhu, Guangtun},
  year = 2016,
  month = dec,
  number = {arXiv:1612.06037},
  eprint = {1612.06037},
  primaryclass = {astro-ph},
  publisher = {arXiv},
  url = {http://arxiv.org/abs/1612.06037},
  urldate = {2024-06-08},
  archiveprefix = {arXiv},
  langid = {english}
}

@article{Wang2015,
  title = {{{EVIDENCE FOR PHOTOIONIZATION-DRIVEN BROAD ABSORPTION LINE VARIABILITY}}},
  author = {Wang, Tinggui and Yang, Chenwei and Wang, Huiyuan and Ferland, Gary},
  year = 2015,
  month = dec,
  journal = {The Astrophysical Journal},
  volume = {814},
  number = {2},
  pages = {150},
  issn = {1538-4357},
  doi = {10.1088/0004-637X/814/2/150},
  url = {https://iopscience.iop.org/article/10.1088/0004-637X/814/2/150},
  urldate = {2023-03-27},
  langid = {english}
}

@ARTICLE{Elvis2000,
       author = {{Elvis}, Martin},
        title = "{A Structure for Quasars}",
      journal = {\apj},
         year = 2000,
        month = dec,
       volume = {545},
       number = {1},
        pages = {63-76},
          doi = {10.1086/317778},
archivePrefix = {arXiv},
       eprint = {astro-ph/0008064},
 primaryClass = {astro-ph},
       adsurl = {https://ui.adsabs.harvard.edu/abs/2000ApJ...545...63E}
}

@ARTICLE{Hazard1984,
       author = {{Hazard}, C. and {Morton}, D.~C. and {Terlevich}, R. and {McMahon}, R.},
        title = "{Nine new quasi-stellar objects with broad absorption lines.}",
      journal = {\apj},
         year = 1984,
        month = jul,
       volume = {282},
        pages = {33-52},
          doi = {10.1086/162174},
       adsurl = {https://ui.adsabs.harvard.edu/abs/1984ApJ...282...33H}
}

@article{Zhang2010,
  title = {Low-z {{Mg II Broad Absorption-line Quasars}} from the {{Sloan Digital Sky Survey}}},
  author = {Zhang, Shaohua and Wang, Ting-Gui and Wang, Huiyuan and Zhou, Hongyan and Dong, Xiao-Bo and Wang, Jian-Guo},
  year = 2010,
  month = may,
  journal = {The Astrophysical Journal},
  volume = {714},
  pages = {367--383},
  publisher = {IOP},
  issn = {0004-637X},
  doi = {10.1088/0004-637X/714/1/367},
  url = {https://ui.adsabs.harvard.edu/abs/2010ApJ...714..367Z},
  urldate = {2026-06-06}
}

@ARTICLE{DESIdr1,
       author = {{DESI Collaboration} and {Abdul Karim}, M. and {Adame}, A.~G. and {Aguado}, D. and {Aguilar}, J. and {Ahlen}, S. and {Alam}, S. and {Aldering}, G. and {Alexander}, D.~M. and {Alfarsy}, R. and {Allen}, L. and {Allende Prieto}, C. and {Alves}, O. and {Anand}, A. and {Andrade}, U. and {Armengaud}, E. and {Avila}, S. and {Aviles}, A. and {Awan}, H. and {Bailey}, S. and {Baleato Lizancos}, A. and {Ballester}, O. and {Bault}, A. and {Bautista}, J. and {Bean}, R. and {Behera}, J. and {BenZvi}, S. and {Beraldo e Silva}, L. and {Bermejo-Climent}, J.~R. and {Beutler}, F. and {Bianchi}, D. and {Blake}, C. and {Blum}, R. and {Bolton}, A.~S. and {Bonici}, M. and {Brieden}, S. and {Brodzeller}, A. and {Brooks}, D. and {Buckley-Geer}, E. and {Burtin}, E. and {Bystr{\"o}m}, A. and {Canning}, R. and {Carnero Rosell}, A. and {Carr}, A. and {Carrilho}, P. and {Casas}, L. and {Castander}, F.~J. and {Cereskaite}, R. and {Cervantes-Cota}, J.~L. and {Chaussidon}, E. and {Chaves-Montero}, J. and {Chen}, S. and {Chen}, X. and {Circosta}, C. and {Claybaugh}, T. and {Cole}, S. and {Cooper}, A.~P. and {Cousinou}, M.-C. and {Cuceu}, A. and {Davis}, T.~M. and {Dawson}, K.~S. and {de Belsunce}, R. and {de la Cruz}, R. and {de la Macorra}, A. and {de Mattia}, A. and {Deiosso}, N. and {Della Costa}, J. and {Demina}, R. and {Demirbozan}, U. and {DeRose}, J. and {Dey}, A. and {Dey}, B. and {Ding}, J. and {Ding}, Z. and {Doel}, P. and {Douglass}, K. and {Dowicz}, M. and {Ebina}, H. and {Edelstein}, J. and {Eisenstein}, D.~J. and {Elbers}, W. and {Emas}, N. and {Escoffier}, S. and {Fagrelius}, P. and {Fan}, X. and {Fanning}, K. and {Favole}, G. and {Fawcett}, V.~A. and {Fern{\'a}ndez-Garc{\'\i}a}, E. and {Ferraro}, S. and {Findlay}, N. and {Font-Ribera}, A. and {Forero-Romero}, J.~E. and {Forero-S{\'a}nchez}, D. and {Frenk}, C.~S. and {G{\"a}nsicke}, B.~T. and {Galbany}, L. and {Garc{\'\i}a-Bellido}, J. and {Garcia-Quintero}, C. and {Garrison}, L.~H. and {Gazta{\~n}aga}, E. and {Gil-Mar{\'\i}n}, H. and {Gloudemans}, A. and {Gnedin}, O.~Y. and {Gontcho A Gontcho}, S. and {Gonzalez}, D. and {Gonzalez-Morales}, A.~X. and {Gonzalez-Perez}, V. and {Gordon}, C. and {Graur}, O. and {Green}, D. and {Gruen}, D. and {Gsponer}, R. and {Guandalin}, C. and {Gutierrez}, G. and {Guy}, J. and {Hahn}, C. and {Han}, J.~J. and {Han}, J. and {He}, S. and {Herrera-Alcantar}, H.~K. and {Heydenreich}, S. and {Honscheid}, K. and {Hou}, J. and {Howlett}, C. and {Huterer}, D. and {Ir{\v{s}}i{\v{c}}}, V. and {Ishak}, M. and {Jacques}, A. and {Jiang}, L. and {Jimenez}, J. and {Jing}, Y.~P. and {Joachimi}, B. and {Joudaki}, S. and {Joyce}, R. and {Jullo}, E. and {Juneau}, S. and {Kara{\c{c}}ayl{\i}}, N.~G. and {Karim}, T. and {Kehoe}, R. and {Kent}, S. and {Khederlarian}, A. and {Kirkby}, D. and {Kisner}, T. and {Kitaura}, F.-S. and {Kizhuprakkat}, N. and {Kong}, H. and {Koposov}, S.~E. and {Kremin}, A. and {Krolewski}, A. and {Lahav}, O. and {Lai}, Y. and {Lamman}, C. and {Lan}, T.-W. and {Landriau}, M. and {Lang}, D. and {Lange}, J.~U. and {Lasker}, J. and {Le Goff}, J.~M. and {Le Guillou}, L. and {Leauthaud}, A. and {Levi}, M.~E. and {Li}, S. and {Li}, T.~S. and {Liu}, W. and {Lodha}, K. and {Lokken}, M. and {Luo}, Y. and {Magneville}, C. and {Manera}, M. and {Manser}, C.~J. and {Margala}, D. and {Martini}, P. and {Maus}, M. and {McCullough}, J. and {McDonald}, P. and {Medina}, G.~E. and {Medina-Varela}, L. and {Meisner}, A. and {Mena-Fern{\'a}ndez}, J. and {Menegas}, A. and {Meneses-Rizo}, J. and {Mezcua}, M. and {Miquel}, R. and {Montero-Camacho}, P. and {Moon}, J. and {Moustakas}, J. and {Mu{\~n}oz-Guti{\'e}rrez}, A. and {Mu noz-Santos}, D. and {Myers}, A.~D. and {Myles}, J. and {Nadathur}, S. and {Najita}, J. and {Napolitano}, L. and {Newman}, J.~A. and {Nikakhtar}, F. and {Nikutta}, R. and {Niz}, G. and {Noriega}, H.~E. and {Nugent}, P.},
        title = "{Data Release 1 of the Dark Energy Spectroscopic Instrument}",
      journal = {\aj},
         year = 2026,
        month = may,
       volume = {171},
       number = {5},
          eid = {285},
        pages = {285},
          doi = {10.3847/1538-3881/ae4c43},
archivePrefix = {arXiv},
       eprint = {2503.14745},
 primaryClass = {astro-ph.CO},
       adsurl = {https://ui.adsabs.harvard.edu/abs/2026AJ....171..285D}
}

@ARTICLE{SDSSdr7,
       author = {{Abazajian}, Kevork N. and {Adelman-McCarthy}, Jennifer K. and {Ag{\"u}eros}, Marcel A. and {Allam}, Sahar S. and {Allende Prieto}, Carlos and {An}, Deokkeun and {Anderson}, Kurt S.~J. and {Anderson}, Scott F. and {Annis}, James and {Bahcall}, Neta A. and {Bailer-Jones}, C.~A.~L. and {Barentine}, J.~C. and {Bassett}, Bruce A. and {Becker}, Andrew C. and {Beers}, Timothy C. and {Bell}, Eric F. and {Belokurov}, Vasily and {Berlind}, Andreas A. and {Berman}, Eileen F. and {Bernardi}, Mariangela and {Bickerton}, Steven J. and {Bizyaev}, Dmitry and {Blakeslee}, John P. and {Blanton}, Michael R. and {Bochanski}, John J. and {Boroski}, William N. and {Brewington}, Howard J. and {Brinchmann}, Jarle and {Brinkmann}, J. and {Brunner}, Robert J. and {Budav{\'a}ri}, Tam{\'a}s and {Carey}, Larry N. and {Carliles}, Samuel and {Carr}, Michael A. and {Castander}, Francisco J. and {Cinabro}, David and {Connolly}, A.~J. and {Csabai}, Istv{\'a}n and {Cunha}, Carlos E. and {Czarapata}, Paul C. and {Davenport}, James R.~A. and {de Haas}, Ernst and {Dilday}, Ben and {Doi}, Mamoru and {Eisenstein}, Daniel J. and {Evans}, Michael L. and {Evans}, N.~W. and {Fan}, Xiaohui and {Friedman}, Scott D. and {Frieman}, Joshua A. and {Fukugita}, Masataka and {G{\"a}nsicke}, Boris T. and {Gates}, Evalyn and {Gillespie}, Bruce and {Gilmore}, G. and {Gonzalez}, Belinda and {Gonzalez}, Carlos F. and {Grebel}, Eva K. and {Gunn}, James E. and {Gy{\"o}ry}, Zsuzsanna and {Hall}, Patrick B. and {Harding}, Paul and {Harris}, Frederick H. and {Harvanek}, Michael and {Hawley}, Suzanne L. and {Hayes}, Jeffrey J.~E. and {Heckman}, Timothy M. and {Hendry}, John S. and {Hennessy}, Gregory S. and {Hindsley}, Robert B. and {Hoblitt}, J. and {Hogan}, Craig J. and {Hogg}, David W. and {Holtzman}, Jon A. and {Hyde}, Joseph B. and {Ichikawa}, Shin-ichi and {Ichikawa}, Takashi and {Im}, Myungshin and {Ivezi{\'c}}, {\v{Z}}eljko and {Jester}, Sebastian and {Jiang}, Linhua and {Johnson}, Jennifer A. and {Jorgensen}, Anders M. and {Juri{\'c}}, Mario and {Kent}, Stephen M. and {Kessler}, R. and {Kleinman}, S.~J. and {Knapp}, G.~R. and {Konishi}, Kohki and {Kron}, Richard G. and {Krzesinski}, Jurek and {Kuropatkin}, Nikolay and {Lampeitl}, Hubert and {Lebedeva}, Svetlana and {Lee}, Myung Gyoon and {Lee}, Young Sun and {French Leger}, R. and {L{\'e}pine}, S{\'e}bastien and {Li}, Nolan and {Lima}, Marcos and {Lin}, Huan and {Long}, Daniel C. and {Loomis}, Craig P. and {Loveday}, Jon and {Lupton}, Robert H. and {Magnier}, Eugene and {Malanushenko}, Olena and {Malanushenko}, Viktor and {Mandelbaum}, Rachel and {Margon}, Bruce and {Marriner}, John P. and {Mart{\'\i}nez-Delgado}, David and {Matsubara}, Takahiko and {McGehee}, Peregrine M. and {McKay}, Timothy A. and {Meiksin}, Avery and {Morrison}, Heather L. and {Mullally}, Fergal and {Munn}, Jeffrey A. and {Murphy}, Tara and {Nash}, Thomas and {Nebot}, Ada and {Neilsen}, Jr., Eric H. and {Newberg}, Heidi Jo and {Newman}, Peter R. and {Nichol}, Robert C. and {Nicinski}, Tom and {Nieto-Santisteban}, Maria and {Nitta}, Atsuko and {Okamura}, Sadanori and {Oravetz}, Daniel J. and {Ostriker}, Jeremiah P. and {Owen}, Russell and {Padmanabhan}, Nikhil and {Pan}, Kaike and {Park}, Changbom and {Pauls}, George and {Peoples}, Jr., John and {Percival}, Will J. and {Pier}, Jeffrey R. and {Pope}, Adrian C. and {Pourbaix}, Dimitri and {Price}, Paul A. and {Purger}, Norbert and {Quinn}, Thomas and {Raddick}, M. Jordan and {Re Fiorentin}, Paola and {Richards}, Gordon T. and {Richmond}, Michael W. and {Riess}, Adam G. and {Rix}, Hans-Walter and {Rockosi}, Constance M. and {Sako}, Masao and {Schlegel}, David J. and {Schneider}, Donald P. and {Scholz}, Ralf-Dieter and {Schreiber}, Matthias R. and {Schwope}, Axel D. and {Seljak}, Uro{\v{s}} and {Sesar}, Branimir and {Sheldon}, Erin and {Shimasaku}, Kazu and {Sibley}, Valena C. and {Simmons}, A.~E. and {Sivarani}, Thirupathi and {Allyn Smith}, J. and {Smith}, Martin C. and {Smol{\v{c}}i{\'c}}, Vernesa and {Snedden}, Stephanie A. and {Stebbins}, Albert and {Steinmetz}, Matthias and {Stoughton}, Chris and {Strauss}, Michael A. and {SubbaRao}, Mark and {Suto}, Yasushi and {Szalay}, Alexander S. and {Szapudi}, Istv{\'a}n and {Szkody}, Paula and {Tanaka}, Masayuki and {Tegmark}, Max and {Teodoro}, Luis F.~A. and {Thakar}, Aniruddha R. and {Tremonti}, Christy A. and {Tucker}, Douglas L. and {Uomoto}, Alan and {Vanden Berk}, Daniel E. and {Vandenberg}, Jan and {Vidrih}, S. and {Vogeley}, Michael S. and {Voges}, Wolfgang and {Vogt}, Nicole P. and {Wadadekar}, Yogesh and {Watters}, Shannon and {Weinberg}, David H. and {West}, Andrew A. and {White}, Simon D.~M. and {Wilhite}, Brian C. and {Wonders}, Alainna C. and {Yanny}, Brian and {Yocum}, D.~R.},
        title = "{The Seventh Data Release of the Sloan Digital Sky Survey}",
      journal = {\apjs},
         year = 2009,
        month = jun,
       volume = {182},
       number = {2},
        pages = {543-558},
          doi = {10.1088/0067-0049/182/2/543},
archivePrefix = {arXiv},
       eprint = {0812.0649},
 primaryClass = {astro-ph},
       adsurl = {https://ui.adsabs.harvard.edu/abs/2009ApJS..182..543A}
}

@ARTICLE{BOSS,
       author = {{Smee}, Stephen A. and {Gunn}, James E. and {Uomoto}, Alan and {Roe}, Natalie and {Schlegel}, David and {Rockosi}, Constance M. and {Carr}, Michael A. and {Leger}, French and {Dawson}, Kyle S. and {Olmstead}, Matthew D. and {Brinkmann}, Jon and {Owen}, Russell and {Barkhouser}, Robert H. and {Honscheid}, Klaus and {Harding}, Paul and {Long}, Dan and {Lupton}, Robert H. and {Loomis}, Craig and {Anderson}, Lauren and {Annis}, James and {Bernardi}, Mariangela and {Bhardwaj}, Vaishali and {Bizyaev}, Dmitry and {Bolton}, Adam S. and {Brewington}, Howard and {Briggs}, John W. and {Burles}, Scott and {Burns}, James G. and {Castander}, Francisco Javier and {Connolly}, Andrew and {Davenport}, James R.~A. and {Ebelke}, Garrett and {Epps}, Harland and {Feldman}, Paul D. and {Friedman}, Scott D. and {Frieman}, Joshua and {Heckman}, Timothy and {Hull}, Charles L. and {Knapp}, Gillian R. and {Lawrence}, David M. and {Loveday}, Jon and {Mannery}, Edward J. and {Malanushenko}, Elena and {Malanushenko}, Viktor and {Merrelli}, Aronne James and {Muna}, Demitri and {Newman}, Peter R. and {Nichol}, Robert C. and {Oravetz}, Daniel and {Pan}, Kaike and {Pope}, Adrian C. and {Ricketts}, Paul G. and {Shelden}, Alaina and {Sandford}, Dale and {Siegmund}, Walter and {Simmons}, Audrey and {Smith}, D. Shane and {Snedden}, Stephanie and {Schneider}, Donald P. and {SubbaRao}, Mark and {Tremonti}, Christy and {Waddell}, Patrick and {York}, Donald G.},
        title = "{The Multi-object, Fiber-fed Spectrographs for the Sloan Digital Sky Survey and the Baryon Oscillation Spectroscopic Survey}",
      journal = {\aj},
         year = 2013,
        month = aug,
       volume = {146},
       number = {2},
          eid = {32},
        pages = {32},
          doi = {10.1088/0004-6256/146/2/32},
archivePrefix = {arXiv},
       eprint = {1208.2233},
 primaryClass = {astro-ph.IM},
       adsurl = {https://ui.adsabs.harvard.edu/abs/2013AJ....146...32S}
}

@ARTICLE{BOSS_SDSS3,
       author = {{Dawson}, Kyle S. and {Schlegel}, David J. and {Ahn}, Christopher P. and {Anderson}, Scott F. and {Aubourg}, {\'E}ric and {Bailey}, Stephen and {Barkhouser}, Robert H. and {Bautista}, Julian E. and {Beifiori}, Alessandra and {Berlind}, Andreas A. and {Bhardwaj}, Vaishali and {Bizyaev}, Dmitry and {Blake}, Cullen H. and {Blanton}, Michael R. and {Blomqvist}, Michael and {Bolton}, Adam S. and {Borde}, Arnaud and {Bovy}, Jo and {Brandt}, W.~N. and {Brewington}, Howard and {Brinkmann}, Jon and {Brown}, Peter J. and {Brownstein}, Joel R. and {Bundy}, Kevin and {Busca}, N.~G. and {Carithers}, William and {Carnero}, Aurelio R. and {Carr}, Michael A. and {Chen}, Yanmei and {Comparat}, Johan and {Connolly}, Natalia and {Cope}, Frances and {Croft}, Rupert A.~C. and {Cuesta}, Antonio J. and {da Costa}, Luiz N. and {Davenport}, James R.~A. and {Delubac}, Timoth{\'e}e and {de Putter}, Roland and {Dhital}, Saurav and {Ealet}, Anne and {Ebelke}, Garrett L. and {Eisenstein}, Daniel J. and {Escoffier}, S. and {Fan}, Xiaohui and {Filiz Ak}, N. and {Finley}, Hayley and {Font-Ribera}, Andreu and {G{\'e}nova-Santos}, R. and {Gunn}, James E. and {Guo}, Hong and {Haggard}, Daryl and {Hall}, Patrick B. and {Hamilton}, Jean-Christophe and {Harris}, Ben and {Harris}, David W. and {Ho}, Shirley and {Hogg}, David W. and {Holder}, Diana and {Honscheid}, Klaus and {Huehnerhoff}, Joe and {Jordan}, Beatrice and {Jordan}, Wendell P. and {Kauffmann}, Guinevere and {Kazin}, Eyal A. and {Kirkby}, David and {Klaene}, Mark A. and {Kneib}, Jean-Paul and {Le Goff}, Jean-Marc and {Lee}, Khee-Gan and {Long}, Daniel C. and {Loomis}, Craig P. and {Lundgren}, Britt and {Lupton}, Robert H. and {Maia}, Marcio A.~G. and {Makler}, Martin and {Malanushenko}, Elena and {Malanushenko}, Viktor and {Mandelbaum}, Rachel and {Manera}, Marc and {Maraston}, Claudia and {Margala}, Daniel and {Masters}, Karen L. and {McBride}, Cameron K. and {McDonald}, Patrick and {McGreer}, Ian D. and {McMahon}, Richard G. and {Mena}, Olga and {Miralda-Escud{\'e}}, Jordi and {Montero-Dorta}, Antonio D. and {Montesano}, Francesco and {Muna}, Demitri and {Myers}, Adam D. and {Naugle}, Tracy and {Nichol}, Robert C. and {Noterdaeme}, Pasquier and {Nuza}, Sebasti{\'a}n E. and {Olmstead}, Matthew D. and {Oravetz}, Audrey and {Oravetz}, Daniel J. and {Owen}, Russell and {Padmanabhan}, Nikhil and {Palanque-Delabrouille}, Nathalie and {Pan}, Kaike and {Parejko}, John K. and {P{\^a}ris}, Isabelle and {Percival}, Will J. and {P{\'e}rez-Fournon}, Ismael and {P{\'e}rez-R{\`a}fols}, Ignasi and {Petitjean}, Patrick and {Pfaffenberger}, Robert and {Pforr}, Janine and {Pieri}, Matthew M. and {Prada}, Francisco and {Price-Whelan}, Adrian M. and {Raddick}, M. Jordan and {Rebolo}, Rafael and {Rich}, James and {Richards}, Gordon T. and {Rockosi}, Constance M. and {Roe}, Natalie A. and {Ross}, Ashley J. and {Ross}, Nicholas P. and {Rossi}, Graziano and {Rubi{\~n}o-Martin}, J.~A. and {Samushia}, Lado and {S{\'a}nchez}, Ariel G. and {Sayres}, Conor and {Schmidt}, Sarah J. and {Schneider}, Donald P. and {Sc{\'o}ccola}, C.~G. and {Seo}, Hee-Jong and {Shelden}, Alaina and {Sheldon}, Erin and {Shen}, Yue and {Shu}, Yiping and {Slosar}, An{\v{z}}e and {Smee}, Stephen A. and {Snedden}, Stephanie A. and {Stauffer}, Fritz and {Steele}, Oliver and {Strauss}, Michael A. and {Streblyanska}, Alina and {Suzuki}, Nao and {Swanson}, Molly E.~C. and {Tal}, Tomer and {Tanaka}, Masayuki and {Thomas}, Daniel and {Tinker}, Jeremy L. and {Tojeiro}, Rita and {Tremonti}, Christy A. and {Vargas Maga{\~n}a}, M. and {Verde}, Licia and {Viel}, Matteo and {Wake}, David A. and {Watson}, Mike and {Weaver}, Benjamin A. and {Weinberg}, David H. and {Weiner}, Benjamin J. and {West}, Andrew A. and {White}, Martin and {Wood-Vasey}, W.~M. and {Yeche}, Christophe and {Zehavi}, Idit and {Zhao}, Gong-Bo and {Zheng}, Zheng},
        title = "{The Baryon Oscillation Spectroscopic Survey of SDSS-III}",
      journal = {\aj},
         year = 2013,
        month = jan,
       volume = {145},
       number = {1},
          eid = {10},
        pages = {10},
          doi = {10.1088/0004-6256/145/1/10},
archivePrefix = {arXiv},
       eprint = {1208.0022},
 primaryClass = {astro-ph.CO},
       adsurl = {https://ui.adsabs.harvard.edu/abs/2013AJ....145...10D}
}

@ARTICLE{He2017,
       author = {{He}, Zhicheng and {Wang}, Tinggui and {Zhou}, Hongyan and {Bian}, Weihao and {Liu}, Guilin and {Yang}, Chenwei and {Dou}, Liming and {Sun}, Luming},
        title = "{Variation of Ionizing Continuum: The Main Driver of Broad Absorption Line Variability}",
      journal = {\apjs},
         year = 2017,
        month = apr,
       volume = {229},
       number = {2},
          eid = {22},
        pages = {22},
          doi = {10.3847/1538-4365/aa647a},
archivePrefix = {arXiv},
       eprint = {1703.00716},
 primaryClass = {astro-ph.GA},
       adsurl = {https://ui.adsabs.harvard.edu/abs/2017ApJS..229...22H}
}

@ARTICLE{He2025,
       author = {{He}, Zhicheng and {Wang}, Tinggui},
        title = "{Discovery of the asymmetric effect in the response of photoionization gas}",
      journal = {\aap},
         year = 2025,
        month = nov,
       volume = {703},
          eid = {A305},
        pages = {A305},
          doi = {10.1051/0004-6361/202556191},
archivePrefix = {arXiv},
       eprint = {2311.16463},
 primaryClass = {astro-ph.GA},
       adsurl = {https://ui.adsabs.harvard.edu/abs/2025A&A...703A.305H}
}
\bibliographystyle{aasjournalv7}


\end{document}